\documentclass{ethpaper}
\usepackage{graphicx}
\usepackage{subfigure}
\usepackage{amssymb}
\begin{document}
\begin{titlepage}
\ethnote{}
\title{High-energy proton induced damage study\\
of scintillation light output\\
from PbWO${_4}$ calorimeter crystals}
\begin{Authlist}
P.~Lecomte, D.~Luckey, F.~Nessi-Tedaldi, F.~Pauss
\Instfoot{eth}{Institute for Particle Physics, ETH Zurich, 8093 Z\"urich, Switzerland}
\end{Authlist}
\maketitle

\begin{abstract}

Eight PbWO$_4$ crystals produced for the electromagnetic calorimeter
of the CMS experiment at LHC have been irradiated in a 20\,GeV/c
proton beam up to fluences of $5.4 \times 10^{13}\;\mathrm{p/cm}^{2}$.
The damage recovery in these crystals, stored in the dark at room
temperature, has been followed for over a year. Comparative
irradiations with $^{60}$Co photons have been performed on seven other
crystals using a dose rate of 1\,kGy/h.  The issue whether hadrons
cause a specific damage to the scintillation mechanism has been
studied through light output measurements on the irradiated crystals
using cosmic rays.  The correlation between light output changes and
light transmission changes is measured to be the same for
proton-irradiated crystals and for $\gamma$-irradiated crystals.
Thus, within the precision of the measurements and for the explored
range of proton fluences, no additional, hadron-specific damage to the
scintillation mechanism is observed.
\end{abstract}

\vspace{7cm}
\conference{Submitted to Elsevier Science}

\end{titlepage}

\section{Introduction}
\label{s-int}
The way large hadron fluxes affect $\mathrm{PbWO}_4$ crystals has
become relevant with the construction of LHC~\cite{r-FN1}.  For the
CMS experiment, charged hadron fluences have been
calculated~\cite{r-TRA} for an integrated luminosity of $5\times
10^5\;\; \mathrm{pb}^{-1}$, which nominally corresponds to 10 years of
running at LHC, yielding in the electromagnetic calorimeter (ECAL)
barrel up to $\sim 10^{12}$ and in the end caps $\sim 10^{14}$ charged
hadrons/$\mathrm{cm}^2$.  It had to be ascertained whether such fluxes
cause a specific, possibly cumulative, damage, and if so, what its
quantitative importance is and whether it only affects light
transmission (LT) or also the scintillation mechanism.  A
hadron-specific damage could arise from the production, above a
$\sim$20 MeV threshold, of heavy fragments with up to 10 $\mu$m range
and energies up to $\sim$100 MeV, causing a displacement of lattice
atoms and energy losses, along their path, up to 50 000 times the one of
minimum-ionising particles. The damage caused by these processes is
likely to be different from the one caused by ionising radiation and
could be cumulative.

In a study we have published recently~\cite{r-ETH2}, the primarily
investigated quantity was the damage to light transmission, because it
can be measured very accurately. Furthermore, all damage observed in
earlier tests ~\cite{r-BAT1,r-AUF} could always be ascribed to the
ionising dose associated with the hadron flux, apart from some indication
for a hadron-specific damage in BGO, which can be
extracted~\cite{r-FN2} from existing data. None of the previous tests
on lead tungstate was extended however to the full integrated fluences
expected at the LHC.

Eight CMS production crystals of consistent quality were irradiated at
the IRRAD1 facility of the CERN PS accelerator T7 beam
line~\cite{r-GLA} in a 20 GeV/c proton flux of either
$10^{12}\;\mathrm{p\; cm}^{-2}\mathrm{h}^{-1}$ (crystals {\em a, b, c, d, h})
or of $10^{13}\;\mathrm{p\; cm}^{-2}\mathrm{h}^{-1}$ (crystals {\em E, F, G})
\footnote{Labels `prime' (respectively `double-prime') will be used
  after the letter identifying the crystal, in order to indicate a
  second (or third) irradiation of the same crystal.}.  These crystals
have the shape of truncated pyramids with nearly parallelepipedic
dimensions of $2.4 \times 2.4\; \mathrm{cm}^2$ and a length of 23 cm.
The exact fluxes and fluences each crystal was exposed to are listed
in Ref.~\cite{r-ETH2}.  The maximum fluence, of $5.4 \times
10^{13}\;\mathrm{p/cm}^{2}$, was reached for crystal {\em a''}.  To
disentangle the contribution to damage from the associated ionising
dose, complementary $^{60}\mathrm{Co}\; \gamma$-irradiations were
performed at a dose rate of 1 kGy/h on further seven crystals ({\em t,
  u, v, w, x, y, z}) at the ENEA Casaccia Calliope irradiation
facility~\cite{r-BAC}, since a flux of
$10^{12}\;\mathrm{p\; cm}^{-2}\mathrm{h}^{-1}$ has an associated ionising
dose rate in $\mathrm{PbWO}_4$ of 1 kGy/h.  The maximum dose of $50.3$
kGy was reached for crystal {\em v}, just a factor $\sim 2$ below the
dose reached in proton irradiations, which for {\em a''} was 97.5 kGy.

The study of transmission damage revealed that proton irradiation
decreases the light transmission for all wavelengths and moves the
ultra-violet LT band-edge to longer wavelengths, as can be seen in
Fig. 9 of Ref.~\cite{r-ETH2}.  In $\gamma$-irradiated crystals, the
band-edge does not shift at all, even after the highest cumulated
dose: only the well-known absorption band~\cite{r-ZHU1} appears around
420 nm. These data demonstrate the qualitatively different nature of
proton-induced and $\gamma$-induced damage.  The correlation we found,
between the induced absorption coefficient at the peak of
scintillation-emission wavelength, $\mu_{IND}\mathrm{(420\;
  nm)}$~\cite{r-ETH2}, and fluence, is consistent with a linear
behaviour over two orders of magnitude, showing that proton-induced
damage in $\mathrm{PbWO}_4$ is predominantly cumulative, unlike
$\gamma$-induced damage, which reaches
equilibrium~\cite{r-ZHU1,r-ETH1}.  The data for $\mu_{IND}$ versus
light wavelength for proton-irradiated crystals show a $\lambda^{-4}$
dependence, which is absent in $\gamma$ - irradiated crystals.  This
is an indication of Rayleigh scattering from small centres of severe
damage, as it might be caused by the high energy deposition of heavily
ionising fragments along their path, locally changing the crystal
structure. Taking into account the difference in composition and
energy spectra between 20 GeV/c protons and CMS, simulations indicate
that the test results cover the CMS running conditions up to a
pseudorapidity of $\sim 2.6$. An experimental confirmation is expected
from a planned pion-irradiation of $\mathrm{PbWO}_4$, closer to the
CMS particle spectrum average~\cite{r-ETH2}.

However, the relevant quantity for detector operation is the
scintillation light output (LO). Detector calibration through a light
injection system, as foreseen by CMS~~\cite{r-TDR}, can only work,
as long as changes in light output are all due to changes in light
transmission. This had been verified extensively so far for ionising
radiation, but for high hadron fluxes it still had to be established:
the present work deals with this question. In the following, the
measurement setup and technique are described, the data analysis is
explained, and results are given, in comparison, for proton-irradiated
crystals and for $\gamma$ -irradiated crystals.
\section{The setup}
\label{s-set}
Light output from a crystal is commonly measured in the laboratory
using photons from a $^{60}\mathrm{Co}$ or $^{137}\mathrm{Cs}$ source
to excite scintillation.  The light output is then determined from the
position, in the spectrum, of the photoelectric peak produced by 
photons which deposit all their energy in the crystal. For lead
tungstate crystals, the light output is modest, $\sim 10$
photoelectrons/MeV of incoming particle energy, as detected on the
bialkali photocathode of a photomultiplier (PM) covering the whole
back face of the crystal. Thus, the photoelectric peak is broadened by
statistical fluctuations and merged with the Compton edge.  In the
present study, the hadron-irradiated crystals will remain radioactive
for years, and thus act as sources for $\gamma$ and $\beta$ emission
over an extended range of energies, peaked at the lowest detectable
values and tailing off towards larger ones.  The photoelectric peak
from any common laboratory $\gamma$ source is swamped by this
spectrum, and a different method of LO determination had to be
adopted.

As an alternative, cosmic muons passing a crystal transversely were
used.  These muons deposit 24.5 MeV in the crystal, an energy deposit
which is well above all $\gamma$ energies emitted by radioactive
isotopes. Using these muons to excite scintillation gave us a light
output well above the crystal's own background.

An individual crystal was placed in horizontal position, with plastic
scintillator counters above and below, to select nearly vertical
muons.  Since only a relative measurement of LO loss was sought, large
rectangular plastic scintillators were used, 24 cm long and 23 cm
wide, sitting at 22 cm distance above and below the crystal. These
counters yielded a large angular acceptance, allowing for muons
traversing the crystal at angles between $0^o$ and $37^o$ from the
vertical, leaving thus an energy deposit $E_{dep}$ between 24.5 MeV
and 30.6 MeV.
\section{Measurements and systematics}
\label{s-mea}
\begin{figure}[ht]
\begin{center}\footnotesize
\mbox{\includegraphics[width=12cm,viewport= 0 0 565 394, clip=true]
{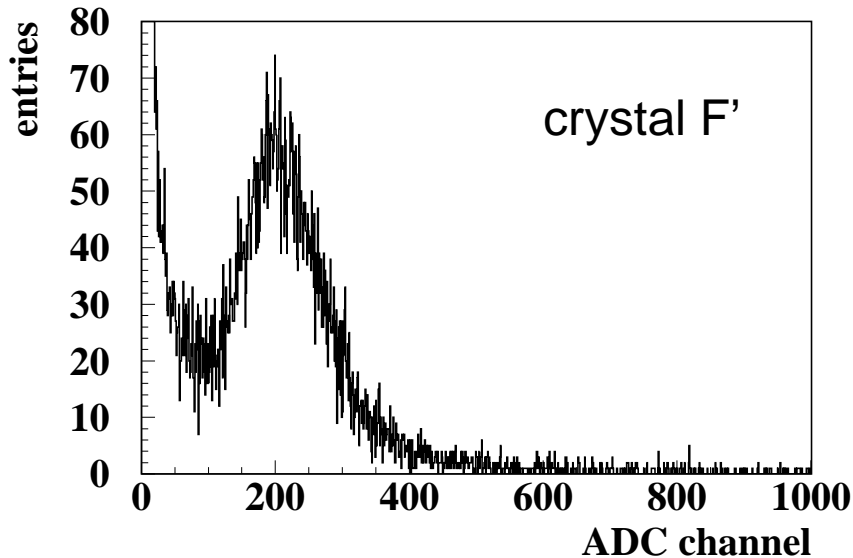}}
\end{center}
\caption{Spectrum for crystal {\em F'}.
 \label{f-Fp}}
\end{figure}
For LO measurements, each crystal was wrapped in
Tyvek\texttrademark~\cite{r-TYV} on all sides but its large end, which
was coupled via an air gap to a Philips XP2262B bialkali-photocathode
PM equipped with a standard CERN base type 4238.  A PVC centering tube
was used for reproducible positioning. The photomultiplier was
operated at a gain of $\sim{10}^8$. The current pulses from its anode
were fed into an Analog-to-Digital converter 2249W from LeCroy, which
has a sensitivity of 0.25 pC. Data acquisition was performed with a PC
running a Labview 7.0 virtual instrument, interfaced via a HYTEC 1331
controller~\cite{r-HYT} to a CAMAC crate. The data acquisition was
triggered by a coincidence of the signals in both plastic scintillator
counters. A CAMAC-controlled Programmable Dual Gate and Delay
Generator LeCroy 2323A defined a 200 ns ADC charge integration
time.  The acquisition of a spectrum with sufficient
statistics for one LO measurement typically took 24 h, yielding
approximately 8000 events in the muon energy deposition peak. The
ambient temperature was recorded at the beginning and at the end of
each measurement, since the lead tungstate LO changes by
$-2\%/^o$C~\cite{r-DAF,r-TDR} at room temperature, and the light output
determined for each run was corrected offline to its value at
$18^o$C. The uncertainty on the correction, coming from possible,
undetected fluctuations of temperature in the crystal over the
duration of a run, amounts to 0.4\% and is negligible compared to the
other error sources affecting the LO determination, as discussed
below. The anode signal was attenuated at the input, by typically 24
db, for the acquisition of a muon spectrum.  Keeping the same PM high
voltage, the stability of the setup, from the PM multiplication chain
to the ADC, was monitored by periodically acquiring the spectrum of
single photoelectrons thermally emitted by the photocathode. In this
case, the anode signal from the PM coupled to the crystal was used for
self-triggering, with no attenuation before the ADC input, and the
discriminator threshold was set low enough to see the valley below the
single photoelectron peak, according to the method we described in
Ref.~\cite{r-ETH1}.  Pedestal data were acquired at the beginning of
every run, and subtracted on a run-by-run basis.  Since we could not
exclude a priori that the crystal radioactivity would affect the
detection efficiency of the setup over time, we also periodically
monitored the stability of the setup by acquiring the cosmic muons
spectrum from a non-irradiated reference crystal, and, when necessary,
corrected all data for relative gain changes.  This procedure allowed
to monitor the whole light detection chain, including the photocathode
quantum efficiency.

As an example, Figure~\ref{f-Fp} shows a cosmic muons spectrum for
{\em F'} after an irradiation with $9.83 \times
10^{12}\;\mathrm{p/cm}^{2}$, where the crystal experienced a
considerable damage in light transmission, corresponding to an induced
absorption coefficient of $\sim 3\; \mathrm{m}^{-1}$. The sharp rise
which is clearly visible below channel 100 is due to light produced by
the radioactivity of the crystal.
\section{Results}
\label{s-res}
\begin{figure}[th]
\begin{center}\footnotesize
\mbox{\includegraphics[width=12cm]{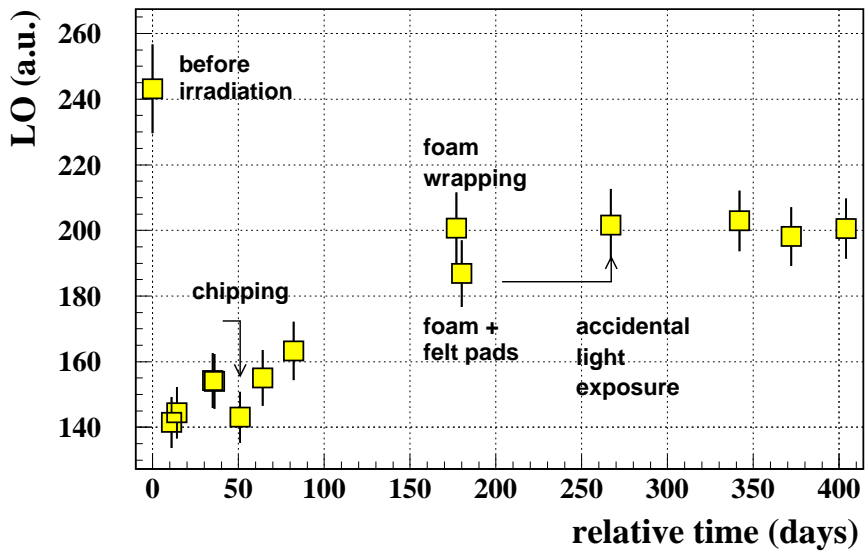}}
\end{center}
\caption{Light output change over time for crystal {\em G}, where the
  long-term, natural room-temperature recovery of damage can be
  observed, along with some accidental changes (see text).
 \label{f-G}}
\end{figure}
\begin{figure}[ht]
\begin{center}\footnotesize
\mbox{\includegraphics[width=12cm]{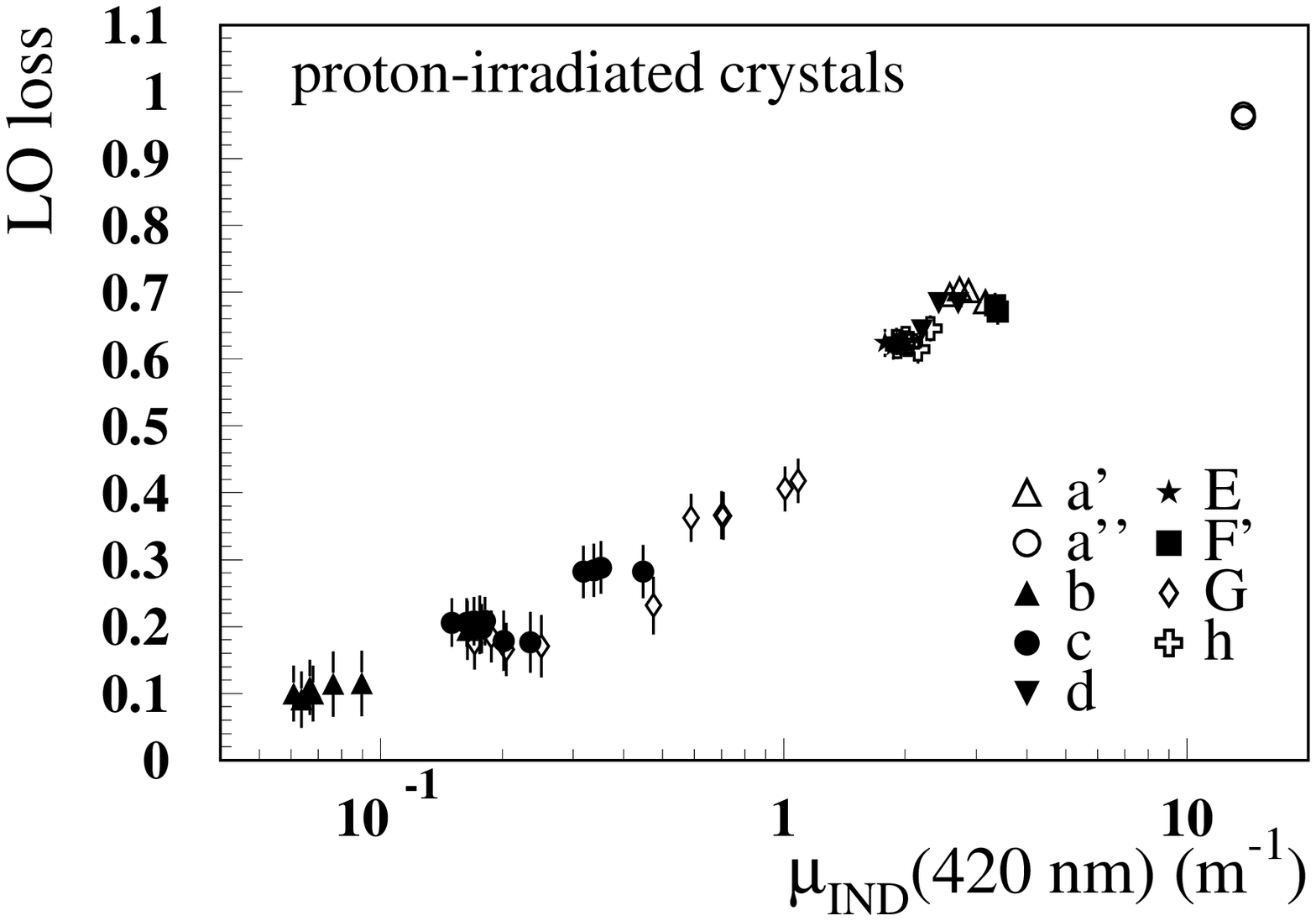}}\\
\mbox{\includegraphics[width=12cm]{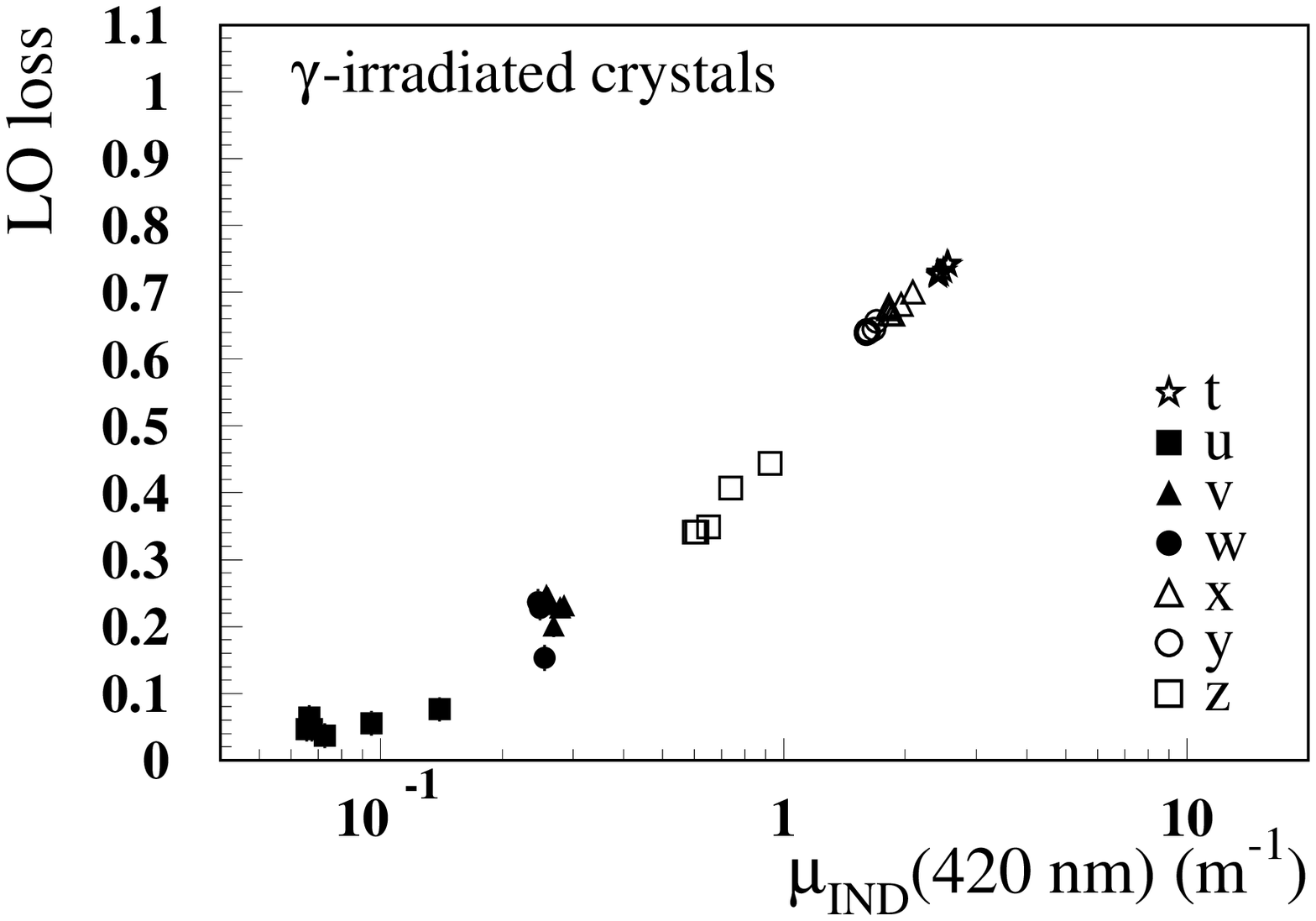}}
\end{center}
\caption{Correlation between $\mu_{IND}\mathrm{(420\; nm)}$ and LO
  loss for proton-induced (top) and $\gamma$-induced (bottom)
  damage. See the text for the systematic scale uncertainty on the
  data (not shown).
 \label{f-LYcorrel}}
\end{figure}
Great care had to be given to sources of systematic errors in the
measurements.  For precise relative LO loss measurements, the
geometrical acceptance of each crystal for triggered muons was kept
constant from one measurement to the next, to ensure keeping the
average muon energy deposit, $< E_{dep}>$, always at the same
value. This was achieved by placing each crystal every time in the
same position and orientation in the setup.  The remaining uncertainty
has been evaluated from the fluctuation of the various LO measurements
for the reference crystal. This was done separately for the period
where proton-irradiated crystals were measured, whose irradiations
were performed between May and October 2003, and for the period where
$\gamma$-irradiated crystals were monitored, which were exposed to
protons in May 2004 and measured over the following 7
months. Reproducibility of positioning was significantly improved over
time as a result of the learning process. The corresponding, relative
uncertainty on LO measurements was determined as the $\sigma$ of the
distribution of values for the reference crystal. For
$\gamma$-irradiated crystals it amounts to 1.7\% and for
proton-irradiated ones, to is 2.4\%, except on the data before
irradiation for crystals {\em a} and {\em F}, which were taken before
monitoring the system stability with frequent reference crystal
measurements, and where it amounts to 3.8\%.  This uncertainty affects
every LO measurement, and it is thus added in quadrature to the
statistical errors. However, the LO change, which we want to correlate
with the damage to light transmission, is determined as the ratio
between light output at a given time after irradiation and the light
output before irradiation.  The uncertainty on the light output
measurement before irradiation thus affects, as an overall scale
uncertainty, the whole set of data for each crystal.

Direct contact was initially used between crystal and PM face, but it
resulted, for the reference crystal and for the proton-irradiated
ones, in the chipping of a few corners, due to the repeated
positioning of the crystals in the setup. This caused a small, but
noticeable loss of collected light with respect to the natural,
room-temperature damage recovery (see Fig.~\ref{f-G}), since the
surface of the large face was slightly reduced by the accident.  To
reduce the risk of chipping, a sheet of polyethylene foam was thus
first wrapped around the Tyvek, to protect the long chamfers.  This
resulted in a small rise of LO signal (see Fig.~\ref{f-G}) because of
the slight change of crystal positioning on the photocathode, whose
efficiency is position-dependent.  Further, four tiny felt pads were
applied, as a protection, to the corners of the big end face, masking
6\% of it.  The LO decreased slightly this way, due to the reduced
area for light collection (Fig.~\ref{f-G}).  Since the amount of
chipping was different from crystal to crystal, and since the changes
cannot be necessarily disentangled from damage recovery, no correction
was applied to the data for these systematic shifts, but we include a
relative systematic uncertainty of 4\% from this effect on the LO
measurements for p-irradiated crystals. The crystals irradiated with
photons are not affected by this uncertainty, since the protection was
adopted before the reference data taking prior to
irradiation. Furthermore, for crystal G one should notice in
Fig.~\ref{f-G} the LO difference between the value at 180 days and the
one at 270 days after irradiation: this rise is due to a step in the
damage recovery, as it can be appreciated in the corresponding LT
recovery data of Fig.11 in Ref.~\cite{r-ETH2}, which originated from
an accidental, prolonged exposure of the crystal to fluorescent room
light.

The correlation between $\mu_{IND}\mathrm{(420\; nm)}$ and LO loss is
shown at the top of Fig.~\ref{f-LYcorrel} for all proton-irradiated
crystals. One should notice the very important damage for {\em a''},
which reached $\mu _{IND}(420\; {\mathrm nm}) \simeq 15\;\mathrm{
  m}^{-1}$ after a fluence of $5.4 \times 10^{13}\;\mathrm{p/cm}^{2}$
and experienced a corresponding LO loss of nearly 96\%.  The
correlation for all $\gamma$-irradiated crystals is shown in the
bottom part of Fig.~\ref{f-LYcorrel}.  The overall, relative,
systematic scale uncertainty affecting the results amounts to 1.7\%
for $\gamma$-irradiated crystals and to 4.7\% for p-irradiated ones,
except for crystals {\em a} and {\em F}, where it is 5.5\%.

Within the precision of our measurements, a clear correlation is
observed between loss of light transmission, expressed through
$\mu_{IND}\mathrm{(420\; nm)}$, and the relative LO loss, both, for
hadron-irradiated crystals and for $\gamma$ irradiated ones.
Furthermore, the two correlations are compatible within the precision
of our measurements. Thus, in the range of proton fluxes and fluences
used for the present irradiations, no hadron-specific alteration to
the crystal scintillation properties has been observed.
\section{Conclusions}
\label{s-con}
A set of lead tungstate crystals of CMS production quality have been
exposed to 20 GeV/c proton fluxes up to $5.4 \times 10^{13}\;
\mathrm{p/cm}^{2}$. For comparison, a different set of similar quality
lead tungstate crystals have been exposed to an equivalent amount of
$^{60}\mathrm{Co}\; \gamma$-radiation in terms of ionising dose rate
and dose.  Measurements of changes in scintillation light output have
been performed and correlated with light transmission changes. Within
the precision of our measurements, the correlations for
proton-irradiated and for $\gamma$-irradiated crystals are
compatible. Thus, for the crystals tested, and within the range of
particle fluxes and fluences considered, we do not observe any
additional, hadron-specific alteration of lead tungstate scintillation
properties.
\section*{Acknowledgments}
We are indebted to R. Steerenberg, who provided us with the required
PS beam conditions for the proton irradiations. We are deeply grateful
to M. Glaser and F. Ravotti, who helped us in operating the
irradiation and dosimetry facilities. The support and practical help
of S. Baccaro and A. Cecilia during the $\gamma$-irradiations at
ENEA-Casaccia were essential and are gratefully acknowledged, as is the
contribution of E.\,Auffray who provided us with the crystals after
having characterised their $\gamma$-radiation hardness at the Geneva
Hospital. Mika Huhtinen contributed in an invaluable way~\cite{r-ETH2}
to the crystal irradiations and to the work on LT damage.

\end{document}